# Entanglement Engineering by Transmon Qubit in a Circuit QED

Ahmad Salmanogli

Çankaya University, Engineering faculty, Electrical and Electronic Department, Ankara, Turkey

**Abstract:** this study significantly emphasizes on the entanglement engineering using a transmon qubit. A transmon qubit is created with two superconducting islands coupled with two Josephson Junction embedded into a transmission line. The transmon qubit energies are manipulated through its coupling to the transmission line. The key factor here is the coupling factor between transmission line and qubit by which the quantum features of the system such as transmon decay rate, energy dispersion, and related coherence time are controlled. To complete knowledge about the design, the system is quantum mechanically analyzed and the related Hamiltonian is derived. Accordingly, the dynamics equation of motions is derived and so the energy dispersion and the coupled system coherence time are investigated. The system engineering should be established in such a way that satisfies the energy dispersion and the coherence time. However, to analyze the entanglement between modes, it needs to calculate the number of photons of the transmission lines and the transmon qubit, and also the phase sensitive cross-correlation. The important section of this study emphasizes on engineering the coupling between the transmon qubit and transmission line to enhance the entanglement. The results show that around the Josephson Junction location where the more coupling is established the more entanglement between modes is created.

## Introduction

Charge qubit traditionally operates in the regime where the coupling energy ($E_J$) is less than the charging energy ($E_c$) [1-3]. This makes the charge qubit so sensitive to the change of charge. It is clear that this type of the qubit can rarely be used in the entanglement applications. It is because the noise has a critical effect on the mentioned properties; therefore, the qubit candidate to be used in the entanglement applications should be designed in such a way to severely subside the noise effect. There are different types of qubit [2-8] such as phase qubit, flux qubit, and transmon qubit. The latter one has been deliberately designed to confine the charge noise effect [1, 2]. The transmon qubit is commonly operated at the strong coupling regime. The transmon qubit structure consists of two superconductors islands coupled through two Josephson Junction (JJ) [3-5]. Actually, the shunting connection of the superconductors in the transmon qubit by which a large capacitance is created is the crucial difference between a transmon qubit and charge qubit. Finally, the transmon qubit is embedded into a transmission line to establish a circuit QED. Circuit QED has been designed to enhance the interaction between the incident light and qubit. One of the main advantages of the circuit QED contributes to its flexible definition of the Hamiltonian. Additionally, one can put many qubits into a transmission line to make them all interact with the same bosonic mode [4, 5]. In the circuit QED the embedding qubits play the role of artificial atoms and interact with the microwave photons. Therefore, to avoid the problems associated with inducing noises, the transmon qubit is selected to generate the microwave entangled photon [8-11]. It is because this type of qubit not only produces some useful degrees of freedom due to the coupling of qubit to a transmission line to engineer the system quantum properties, but also is designed in such a way to suppress the charge noise effect.

This study essentially emphasizes on the generation of the entanglement using circuit QED containing the transmon qubit embedding into a transmission line. Entanglement has been used at different applications such as quantum sensory and quantum radar to enhance the system performance [11,19], and quantum imaging to improve the sensitivity [9, 10]. However, it should not be forgotten that the entanglement as a well-known quantum property [9-18] is easily affected by any applied noise and can be easily destroyed.

With knowledge of the above mentioned points, this study initiates with an idea that using a transmon qubit embedding into a transmission line leads to establishing the entangled photons. The main reason behind associates to the nonlinearity property induced by JJ [9, 10] through which the entangled photons are created. For example, two optical photons become entangled via the interaction of a high intensity laser with a nonlinear material [9, 10]. Additionally, there are several traditional approaches to create the entanglement including electro-opto-mechanical converter [11-13], optoelectronic converter [14,15], Josephson junction [16], plasmonic properties [17], and pure electronics circuit (LNA) [18]. Except methods mentioned in [9], [16] and [18], other approaches employ the optical and microwave cavity coupling to generate the entanglement. Cavities coupling introduce some drawbacks through which the entanglement can be easily leaking away. This fact causes the system to operate with a low efficiency. With knowledge of the studies in [9] and [16], in this work in the same way, we want to create the entanglement between modes using the transmon qubit and the nonlinearity induced by JJ. The nonlinearity in a transmon qubit is arisen due to the Cooper pair tunneling in the structure. The Cooper pair can be imagined as a quantum particle (two fermions with opposite spins and equal and opposite momentum) with mass ($2m_e$), where $m_e$ is the free electron mass. The tunneling of the Copper pairs can be imagined due to the wave behavior of the quantum particle that its wave-function penetrates some distance through the barrier between the superconductors in the structure. One of the interesting points about the Cooper pair tunneling is that this phenomenon may still be established by irradiating the junction with microwave radiation. Accordingly, in the present design, the related irradiating (microwave radiation) is created due to the transmission line some modes (not all) coupling to the transmon qubit in which the energy levels are quantized.

For this reason, a system containing a transmon qubit embedding into a transmission line is designed and then one can engineer the structure to manipulate the entanglement between modes. Engineering the structure leads to varying coupling between the transmission line and qubit modes. This factor as a critical factor will be used to engineer the different parameters in the system such as the decay rate and the energy levels in the transmon qubit. In the following, the designed system and its ingredients will be initially explored and then the system will be quantum mechanically analyzed.

**Theory and Background**

*Structure description (transmon qubit coupling with transmission line)*

The schematic of the system is illustrated in Fig. 1 in which two superconductors coupled with a JJ, called transmon qubit, are embedded into a transmission line. Such an illustrated system can hold the standing waves and make them interact with the localized transmon qubit in circuit QED. Since every electric circuit obeys Maxwell's equations working in the classical regime in which the number of the particles are macroscopic and also the temperature is high. In contrast, when the temperature is dramatically decreased around absolute zero the collective degree of freedom of the circuit obeys the Schrödinger equation.

In the structure proposed, the transmission line with a length equal to $\lambda_{inc}$ is driven with a radio-frequency (RF) wave through coupling capacitor $C_{in}$. The generated standing wave across the transmission line is capacitively ($C_g$) coupled to the interdigitated like structure (superconductors) by which a capacitor ($C_B$) is generated to control the coupling energy of the transmon. In a transmon qubit to create the ultra-coupling ($E_J >> E_c$), the charging energy in qubit ($E_c$) is decreased rather than the direct increase of the JJ tunneling energy $E_J$. Accordingly, to increase the transmission line energy coupling to the transmon qubit one can manipulate $C_g$ through which the coupling factor between the transmission line and transmon qubit is varied. Also, $C_J$ is the JJ related capacitance. Additionally, the schematic illustrated in Fig. 1 shows the coupled (blue) and uncoupled (black) modes between the transmission line and transmon qubit. The latter point mentions that all of the transmission modes cannot couple to the transmon qubit. This point will be discussed later. The structure is stretched from $x = -l$ to $x = l$ ($2l = \lambda_{inc}$) and the transmon qubit is placed between $x = -X_j^-$ to $x = X_j^+$. In Fig. 1 all of the structure related parameters as the inset exaggerated figures are shown. In the inset figures the qubit structure containing two superconductors, an insulator between them, and also the coupling of qubit to transmission line are schematically clearly shown.

*Quantum Hamiltonian*

In this section, the designed structure is quantum mechanically analyzed. The standard method to describe the quantum behavior of an electric circuit includes finding the classical Hamiltonian of the model and then imposing the canonical commutation relation on its degree of freedom. The usual degrees of freedom in the classical regime are the voltage in node and current in loop, while for a quantum circuit it is preferred to use node flux and loop charge which are related to the voltage and current, respectively. Regarding to the model depicted in Fig. 1, the continuous form of the Lagrangian is given by:

$$L_c = \frac{1}{2}\int_{-l}^{X_j^-} dx\left\{C_0(\partial_t\varphi(x,t))^2 - \frac{1}{L_0}(\partial_x\varphi(x,t))^2\right\} + \frac{C_g}{2}\left\{(\partial_t\varphi(X_j^-,t)-\partial_t\varphi_J)^2 + (\partial_t\varphi_J - \partial_t\varphi(X_j^+,t))^2\right\}$$
$$\frac{1}{2}\int_{X_j^+}^{+l} dx\left\{C_0(\partial_t\varphi(x,t))^2 - \frac{1}{L_0}(\partial_x\varphi(x,t))^2\right\} + \frac{C_J+C_B}{2}(\partial_t\varphi_J)^2 + E_J\cos(\frac{2\pi\varphi_J}{\varphi_{J0}})$$
(1)

where $\varphi(x,t)$, $\varphi_J$, $C_0$, and $L_0$ are the continuous flux variable along the transmission line, JJ flux, transmission line capacitance (F/m) and inductance (L/m), respectively. In this equation, $\partial_t$ and $\partial_x$ stand for the first derivative respect to the time and coordinate, respectively.

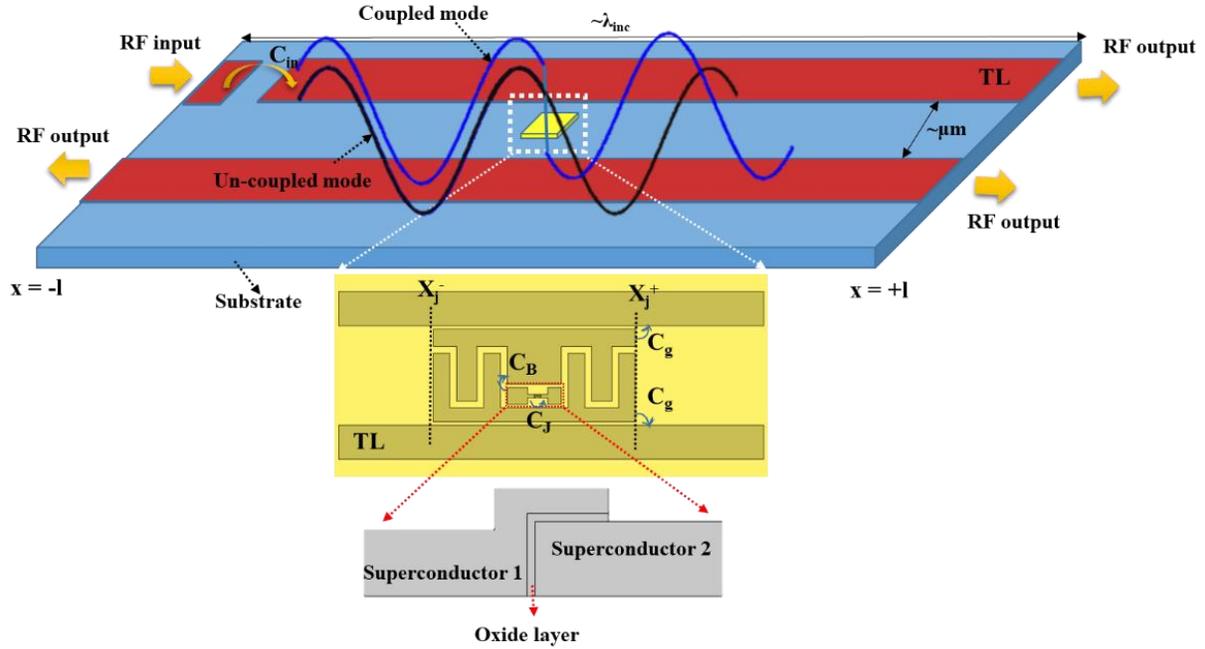

Fig. 1 Schematic of the proposed structure containing transmission lines transferring microwave a long with; embedded transmon qubit at $X_j$ interacting with the transmission line modes; simple charge qubit consisting two superconductor and oxide layer between them.

The last term is Eq. 1 is the energy due to the Cooper pair tunneling through the barrier between the two superconductors. In other words, that is the kinetic energy of the charge carriers moving through the junction. After some algebra and applying some assumptions, the Lagrangian expressed in Eq. 1 can be simplified as:

$$L_c = \frac{1}{2}\int_{-l}^{+l} dx\left\{C_0(\partial_t\varphi(x,t))^2 - \frac{1}{L_0}(\partial_x\varphi(x,t))^2\right\} + \frac{C_g}{2}(\partial_t\varphi(X_j,t)-\partial_t\varphi_J)^2 + \frac{C_J+C_B}{2}(\partial_t\varphi_J)^2 + E_J\cos(\frac{2\pi\varphi_J}{\varphi_{J0}}) \quad (2)$$

Using the orthogonality conditions [2, 4] as $<u_m(x)u_n(x)> = C_\Sigma \delta_{nm}$ and $<\partial_x u_m(x)\partial_x u_n(x)> = C_\Sigma \omega_m^2 \delta_{nm}$, where $C_\Sigma$ is the total capacitance in the structure and standing wave angular frequency $\omega_m = 1/\sqrt{(L_0 C_0)}$, the Lagrangian can be written in terms of the transmission line basis mode (Eigen-function). Accordingly, the transmission line flux $\varphi(x,t)$ needs to decompose in terms of traveling modes as $\varphi(x,t) = \sum_m \Psi_m(t) u_m(x)$, where $\mu_m(t)$ and $u_m(x)$ are a sinusoidal time-dependence function and mode envelope (Eigen-function), respectively. Thus, Lagrangian in terms of the transmission line mode basis is given by:

$$L_c = \sum_m \left\{ \frac{C_\Sigma}{2}(\partial_t \psi_m)^2 - \frac{1}{2L_0}(\psi_m)^2 - C_g u_m(X_J)(\partial_t \psi_m)(\partial_t \varphi_J) \right\} + \frac{C_g + C_J + C_B}{2}(\partial_t \varphi_J)^2 + E_J \cos(\frac{2\pi \varphi_J}{\varphi_{J0}}) \quad (3)$$

In order to do some simplifications, following definitions are applied including $C_G = C_g + C_J + C_B$, $\alpha_m = C_g u_m(X_J)/\sqrt{(C_\Sigma C_G)}$, $\Psi_m = \varphi_m \sqrt{(C_\Sigma)}$, and $\varphi_J = \varphi_{J1}\sqrt{(C_G)}$. Applying the later definitions, the Lagrangian of the system is re-presented as:

$$L_c = \sum_m \left\{ \frac{1}{2}(\partial_t \varphi_m)^2 - \frac{\omega_m^2}{2}(\varphi_m)^2 - \alpha_m(\partial_t \varphi_m)(\partial_t \varphi_{J1}) \right\} + \frac{C_G}{2}(\partial_t \varphi_{J1})^2 + E_J \cos(\frac{2\pi \varphi_{J1}}{\varphi_{J0}}) \quad (4)$$

After calculation of the momentum conjugate variable of the fluxes ($Q_m$ and $Q_J$ are the momentum conjugate variable of the transmission line and transmon qubit, respectively), using the Legendre transformation, and undo the latter changing variable ($\varphi_m \to \Psi_m$, $\varphi_{J1} \to \varphi_J$, and $q_J = Q_J \sqrt{(C_G)}$), the classical Hamiltonian of the system given by:

$$H_c = \sum_m \left\{ \frac{1}{2}\psi_m^2 + \frac{\omega_m^2}{2}Q_m^2 + \frac{\omega_m \alpha_m}{(1-\alpha_s^2)\sqrt{C_G}}(Q_m q_J) + \frac{1}{2C_G(1-\alpha_s^2)}(q_J + q_{DC})^2 \right\} - E_J \cos(\frac{2\pi \varphi_J}{\varphi_{J0}}) \quad (5)$$

In this equation, $q_{DC}$ is the DC part of the $\sum_m(\alpha_m Q_m)$ and the third term relates to the small fluctuation around the DC point. The final step includes applying the canonical quantization method on the classical degree of freedom $[\varphi_k, Q_k] = i\hbar$, to establishing the quantum Hamiltonian of the proposed structure in terms of the ladder operators which is given by:

$$H_q = \sum_m \hbar \omega_m \left( a_m^+ a_m + \frac{1}{2} \right) + \sum_{m,n} \left\{ \left( 2e\beta_m \sqrt{\frac{\hbar}{2\omega_m}} \right)(a_m - a_m^+) \frac{1}{\sqrt{2}} \sqrt[4]{\frac{E_J}{8E_c}}(b_n - b_n^+) \right\}$$

$$+ \sum_{m,n} \sqrt{8E_{cm}E_J}\left( b_n^+ b_n + \frac{1}{2} \right) - \frac{E_{cm}}{12}(b_n + b_n^+)^4 - E_J \quad (6)$$

To express the quantum Hamiltonian in terms of ladder operators, the "cos" function in Eq. 5 needs to expand in the form of the Taylor series. In Eq. 6, $(a_m^+, a_m)$ and $(b_n^+, b_n)$ are the creation and annihilation operators of the transmission line and transmon qubit modes, respectively. Additionally, the remaining parameters are defined as $\beta_m = \alpha_m \omega_m/(1-\alpha_m^2)\sqrt{(C_G)}$ and $E_{cm} = E_c/(1-\alpha_m^2)$, where $E_c$ is the charging energy.

**Results and Discussions**

*Transmon qubit charge dispersion*

In a charge qubit, the sensitivity of the structure to noise can be manipulated by operating a qubit with a specific energy called "sweep spot". That occurred so close to $N_g = 1/2$, where the derivation of energy $\Delta E$ with respect to $N_g$ is zero. $N_g$ is the amount of the charge induced through the external gate voltage. However, in a transmon qubit ($E_J \gg E_c$), the flatten around the sweet spot gets enlarged by the increase of $E_J$. In that situation $\Delta E$ varies less dramatically with unwanted fluctuation of the charge. $\Delta E$ variation is so important because changing the qubit energy gap directly efficiently affects its phase evolution. If $\Delta E$ arose due to the noise and its random behavior, this can gradually disturb the qubit phase coherence. Due to the important points mentioned above, here in this study we tried to examine the different effects on $\Delta E$ and study the effect of the structure that can cause it to vary less $\Delta E$, especially at lower $E_J/E_c$. For this reason, it is necessary to calculate the eigenvalue of the system and examine the energy dispersion for the designed structure. Using the Hamiltonian expressed in Eq. 6 and utilizing Schrödinger equation, one can calculate the energy of the system. For simplicity, the transmission line mode number is truncated to two and the transmon qubit mode numbers is truncated to four. With these assumptions, the energy of the coupling system becomes:

$$E_n = \sum_{m=1}^{2} \hbar \omega_m \left(m + \frac{1}{2}\right) + \sum_{m=1}^{2} \left\{ \left( \sqrt{2} e \beta_m \sqrt[4]{\frac{E_J}{8E_c}} \sqrt{\frac{\hbar}{2\omega_m}} \right) \left( a_m b_n - a_m^+ b_n - a_m b_n^+ + a_m^+ b_n^+ \right) \right\}$$
$$+ \sum_{m=1}^{2} \left\{ \sqrt{8 E_{cm} E_J} \left( n + \frac{1}{2} \right) - \frac{E_{cm}}{12} \left( 6n^2 + 6n + 3 \right) \right\} - E_J \quad (7)$$

The significant point in Eq. 7 is the second term that introduces two different types of interactions; the term ($-a_m b_n^+ - b_n a_m^+$) stands for the beam-splitter-like interaction while the other term ($-a_m b_n - b_n^+ a_m^+$) is used for the amplification-like interaction. The latter one contributes to the interaction between transmission line and transmon qubit modes leading to amplify the number of photons, whereas in the former one the number of photons becomes conserved during the interaction. Through the second term in Eq. 7 (different interaction terms) one can figure out about the coupling the transmission line state to the transmon state; for example, the coupling of the transmission line's first state |1> to the transmon qubit's first state |1> is due to the beam-splitter like interaction, while the coupling of the transmission line's first state |1> to the transmon's third state |3> is due to the amplification like interaction. Similarly, coupling of state |2> from the transmission line to state |2> of the transmon is a beam-splitter like interaction, in contrast coupling of state |2> from the transmission line to state |4> of the transmon is established due to the amplification like interaction.

The other important and critical factor is the effect of $\beta_m$ on the energy levels of the system. In other words, by this factor, one can find which mode of the transmission line can couple to the transmon qubit and which one cannot. For this reason, some related simulations are accomplished and the results are illustrated in Fig. 2. In fact, the different energy levels of the system ($E_n$, n = 1, 2, 3, and 4) derived in Eq. 7 is studied. In this simulation, the JJ is placed at $X_J = 0$. It should be noted that all of the simulations in this study are carried out using constant data from Table. 1.

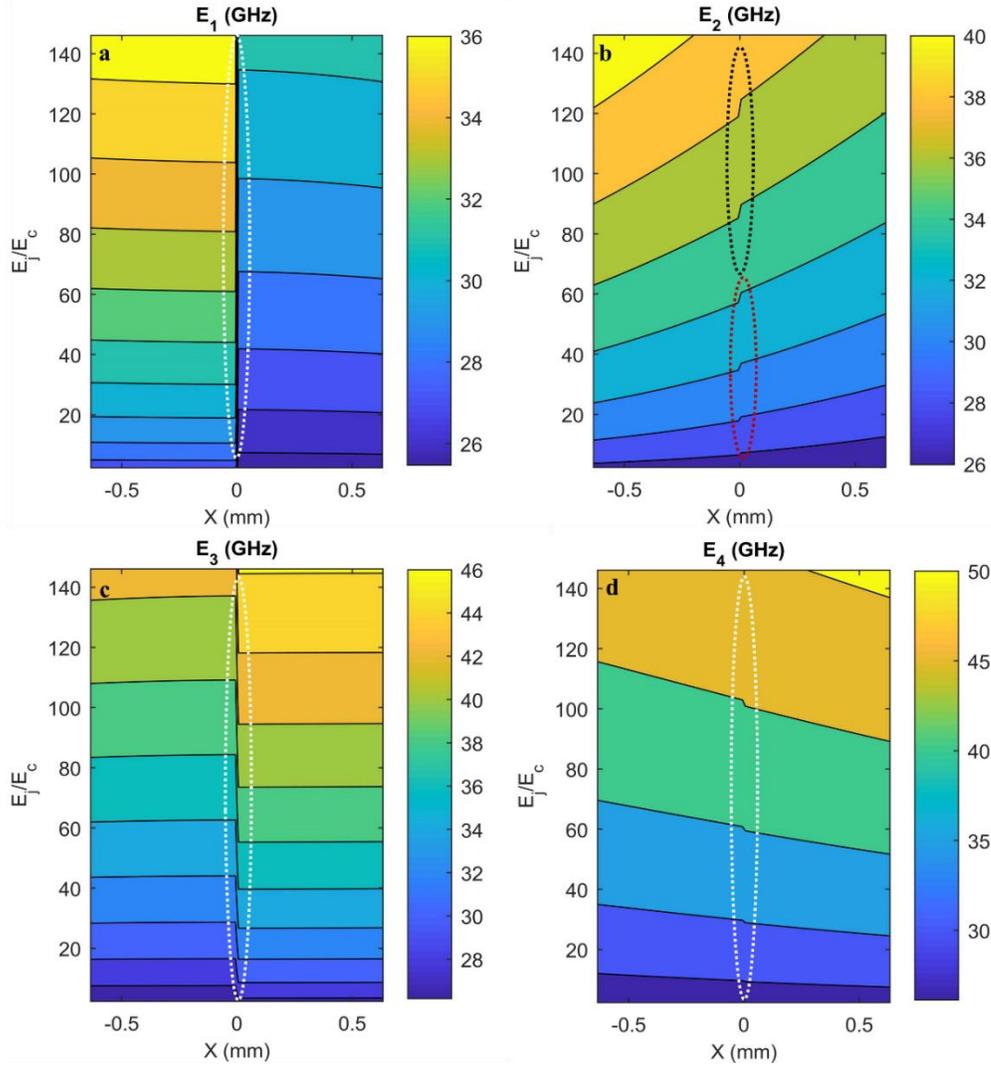

Fig. 2 Energy of the system (GHz) vs. ($E_J/E_c$) and location along the transmission line X (mm); transmon qubit's energies $E_1$ and $E_3$ are affected with coupling of the transmission line odd mode while $E_2$ and $E_4$ are affected with coupling of the transmission line even mode.

The simulation results show that the odd mode of the transmission line is strongly coupled to the transmon qubit while the even mode is merely coupled. This point is indicated with a dashed-elliptic on some sub-figure at the location of JJ. Moreover, the coupling of energy is stronger at low level energy than the higher one; for instance, the coupling of modes at $E_1$ is so stronger than $E_3$ and so forth. This means that the change of the transmon qubit's low level energy is sharper due to the coupling factor. Additionally, the coupling of energy between modes in the system can be manipulated by $E_J/E_c$. For instance, it is shown in Fig. 2b indicated with black and red ellipsoid that the coupling between modes is increased by increasing $E_J/E_c$. The critical factor to manipulate the coupling between modes is $\alpha_m$ which is strongly dependent on $u_m(X_J)$. $u_m(X_J)$ is the Eigen-function (envelope function) of the transmission line at the location of JJ. It is shown that $u_{m=odd}(X_J) \gg u_{m=even}(X_J)$, by which one can easily get the point that there is a negligible overlap between modes when the transmission line modes are even. Additionally, $\Delta E$ is calculated to complete the analysis and the results of the simulation are depicted in Appendix A (Fig. A1). The results show that around the JJ location, $\Delta E$ is sharply changed. This is the point that can dramatically affect the energy dispersion of the transmon. It is attributed to the coupling of the transmission line modes to the qubit embedded in the transmission line. In other words, the odd modes of the transmission line strongly couple to the qubit while the even mode coupling is a slight one. The other important factor to complete the analysis is the anharmonicity created due to the interaction of the modes in the structure which will be investigated in the next part.

*Anharmonicity in the designed system*

The relative anharmonicity as an important factor in the qubit is defined by $\alpha_r = (E_{32} - E_{21})/E_{21}$ and it relates to the minimum pulse duration through $\tau_p \sim 1/|\omega \alpha_r|$, where $\omega$ is the related angular frequency. For a coherent system operating, the minimum pulse duration must remain small compared to the system dephasing times. That is the reason one cannot increase $E_J/E_c$ in an arbitrary way in the design of the qubit.

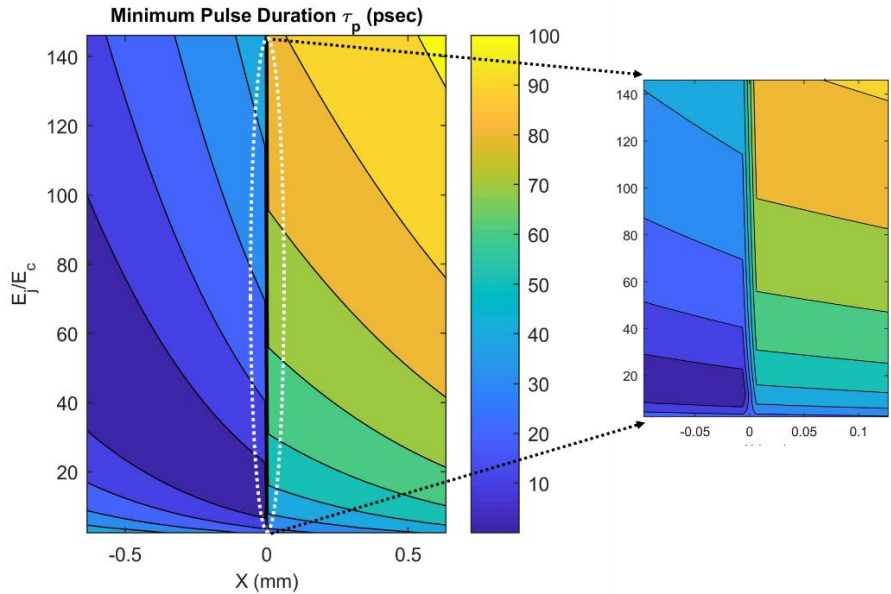

Fig. 3 Minimum pulse duration vs. $E_J/E_c$ and X (mm)

As we expected from the latter illustrated results, the minimum pulse duration would show some odd behavior around the location of the JJ. This is again contributed to the odd mode coupling effect leading to the sharp alteration of $\tau_p$ around X = 0, where the qubit is put there. The simulation result depicted in Fig. 3 reveal that the minimum pulse duration is increased up to 50 psec when $E_J/E_c = 80$ at X ~ $0^+$, which means that the system's dephasing time cannot be less than 50 psec, otherwise system will be operated at a non-coherent mode. Since increasing $\tau_p$ needs to minimize the anharmonicity, this forces the difference between the energy levels of the system to be minimized for example $\omega_{21}$ ~ $\omega_{32}$. The simulation results of $\omega_{21}$ and $\omega_{32}$ are illustrated in Fig. A1 in Appendix A. The equality of the frequencies shown in Fig. A1 means that the detuning frequency of the system ($\omega_{21}$ - $\omega_{32}$) should be either set or fluctuated around zero. This point as a critical factor affecting the entanglement will be discussed later.

Another factor to affect the entanglement between modes in the system is the Purcell factor discussed in the following.

*Transmission line effect on the decay rate of the transmon qubit (Purcell factor)*

The effect of the transmission line on the transmon qubit decay rate is defined as the Purcell factor. This can impact the quantum behavior of the coupled system by which more importantly the entanglement generation by the transmon qubit can be severely affected. To calculate the Purcell factor, it is necessary to derive the dynamic equation of the motion of the system using the total Hamiltonian originally expressed in Eq. 6. Just for the sake of simplicity, the summation operators are ignored and then the simple form of the Hamiltonian is introduced as:

$$H_q = \hbar\omega_m\left(a_m^+ a_m + \frac{1}{2}\right) + \hbar\gamma_m\left(a_m - a_m^+\right)\left(b_n - b_n^+\right) + \hbar\omega_n\left(b_n^+ b_n + \frac{1}{2}\right) - \frac{E_{cm}}{12}\left(b_n + b_n^+\right)^4 - E_J \quad (8)$$

where some definitions are used as $\gamma_m = (e\beta_m/\hbar)(E_J/8E_c)^{0.25}\sqrt{(\hbar/\omega_m)}$ and $\omega_n = \sqrt{(8E_JE_c/\hbar)}$. In this equation, $\gamma_m$ stands for the coupling rate. Using the Langevin equation [21, 22], the system equations of the motion are given by:

$$\begin{aligned}
\dot{a}_m &= -\left(j\Delta_m + \frac{\kappa_m}{2}\right)a_m + j\gamma_m\left(b_n - b_n^+\right) + a_{in} + E_{TL} \\
\dot{b}_n &= -\left(j\Delta_n + \frac{\kappa_n}{2}\right)b_n + j\gamma_m\left(a_m - a_m^+\right) - \frac{E_{cm}}{3j\hbar}\left(b_n + b_n^+\right)^3 + b_{in}
\end{aligned} \quad (9)$$

where $\Delta_m$, $\Delta_n$, $E_{TL}$, $a_{in}$, and $b_{in}$ are the transmission line related detuning frequency, transmon qubit detuning frequency, the transmission line's driven field, transmission line's noise and transmon qubit's noise, respectively. To calculate the Purcell factor [22], we just focus on the transmission line effect on the transmon in the steady state by the calculation of the expectation value of the modes. Thus, the first part of Eq. 9 in the steady state condition becomes $<a_m> = j\gamma_m<(b_n-b_n^+)>/(j\Delta_m+0.5\kappa_m)$. By substituting $<a_m>$ into the second part of Eq. 9 and rearrange the equation in terms of the $b_n$, the transmon qubit new decay rate (or modified decay rate due to the coupling to transmission line) $\kappa_n$' becomes:

$$\dot{b}_n = -\left(j\Delta_n + \frac{\kappa_n}{2}\right)b_n + j\gamma_m 2\operatorname{Im}(a_m) + ... \rightarrow \dot{b}_n = -\left(j\Delta_n + \frac{\kappa_n}{2}\right)b_n - 2\gamma_m \left(\frac{\kappa_m \gamma_m}{2} \frac{b_n}{\Delta_m^2 + \frac{\kappa_m^2}{4}}\right) + ...$$

$$\dot{b}_n = -\left(j\Delta_n + \left\{\frac{\kappa_n}{2} + \frac{4\kappa_m \gamma_m^2}{4\Delta_m^2 + \kappa_m^2}\right\}\right)b_n + ...$$

(10)

where Im stands for the imaginary part. The modified decay rate is affected by the transmission line decay rate, detuning frequency and also strongly by the coupling rate. If one selects $\Delta_m$ so close to zero, $\kappa_n$' = $\kappa_n$/2 + 16$\gamma_m^2$/$\kappa_m$ means that the modified decay rate is strongly affected by the coupling rate ($\gamma_m$). In other words, the transmon qubit's quantum properties are changed due the coupling of the transmission line modes to it. For more information, the contour map of $\kappa_n$' and $\gamma_m$ are depicted in Appendix A in Fig. A2, in which one can clearly observe how and by which rate the mentioned quantities are changed.

Using the dynamic equation of the motion expressed in Eq. 9, one can calculate the transmission line and transmon photon number and also phase sensitive cross-correlation by which the entanglement between modes can be investigated.

*Entanglement induced by the Transmon qubit*

To study about the entanglement being created using a transmon qubit in the system, here in this study the entanglement metric [20] approach is used. In this approach, one needs to calculate the number of output photons of the transmission line and transmon qubit as well as the phase sensitive cross correlation. The latter term is created due to the coupling of the transmon qubit to the transmission line and plays a key role to establish the entanglement in the system. One can start with the dynamics equation of motions expressed in Eq. 9. to calculate the related output photons and also the phase sensitive cross correlation, However, this equation has a nonlinear term and it is better to linearize for the sake of simplicity. To linearize the equations, some variables changing are applied as $a_m = A_m + \delta a_m$ and $b_n = B_n + \delta b_n$ (the capital letters stand for strong field effect and "$\delta$" stands for the fluctuation of the modes around the strong field) and substitute in Eq. 9. The dynamic equation of the fluctuation mode in the form of linearized equations are expressed as:

$$\dot{\delta a}_m = -\left(j\Delta_m + \frac{\kappa_m}{2}\right)\delta a_m + j\gamma_m\left(\delta b_n - \delta b_n^+\right) + \delta a_{in}$$

$$\dot{\delta b}_n = -\left(j\Delta_n + \frac{\kappa_n'}{2}\right)\delta b_n + j\gamma_m\left(\delta a_m - \delta a_m^+\right) + j\gamma_N\left(\delta b_n + \delta b_n^+\right) + \delta b_{in}$$

(11)

where $\gamma_N = E_{cm}*Re(B_n)^2/3\hbar$. In this definition $B_n$ is the transmon qubit strong field and comes from the strong field equations expressed in Appendix B. To calculate the transmission line and transmon photon numbers, Eq. 11 transformed into the Fourier domain and introduced as:

$$\begin{cases} \left(j\Delta_m + \dfrac{\kappa_m}{2}\right)\delta a_m = j\gamma_m\left(\delta b_n - \delta b_n^+\right) + \delta a_{in} \\ \left(j\Delta_n + \dfrac{\kappa_n'}{2}\right)\delta b_n = j\gamma_m\left(\delta a_m - \delta a_m^+\right) + j\gamma_N\left(\delta b_n + \delta b_n^+\right) + \delta b_{in} \end{cases} \quad (12)$$

The transmission line and transmon qubit output photon numbers are calculated in Appendix B and the results given by:

$$\begin{aligned} n_{TL} &= \frac{\gamma_m^2}{|A_0|^2}(2n_T + 1) + \frac{1}{|A_0|^2}n_{ina} \\ n_T &= \frac{\gamma_m^2}{B_1|B_0|^2}(2n_{TL} + 1) + \frac{\gamma_N^2}{B_1|B_0|^2} + \frac{1}{B_1|B_0|^2}n_{inb} \end{aligned} \quad (13)$$

where $n_{ina}$ and $n_{inb}$ are the average number of the photons induced to the system due to the interaction of the transmission line and transmon qubit with the environment, respectively. Solving Eq. 13 gives us the number of transmission line photons $n_{TL}$ and the transmon photon number $n_T$. Finally using the input-output formula ($\delta a_{out\_m} = \delta a_m\sqrt{(2\kappa_m)} - \delta a_{in}$) the related output photon numbers are calculated. Additionally, the phase sensitive cross correlation as an important factor is calculated as:

$$\begin{aligned} d_{mn} &\equiv \langle \delta a_{out\_m}\delta b_{out\_n}\rangle = \langle\left(\delta a_m\sqrt{2\kappa_m} - \delta a_{in}\right)\left(\delta b_n\sqrt{2\kappa_n'} - \delta b_{in}\right)\rangle \\ &= 2\sqrt{\kappa_m\kappa_n'}\langle\delta a_m\delta b_n\rangle \\ &= -2\sqrt{\kappa_m\kappa_n'}\frac{\gamma_m\gamma_N}{A_0B_0}(1+n_T) \end{aligned} \quad (14)$$

After calculation of the related photon numbers and also the phase sensitive cross correlation factor one can utilize entanglement metric [20] to analyze the entanglement between modes:

$$\varepsilon_e = \frac{|d_{nm}|}{\sqrt{n_{oT}n_{oTL}}} \quad (15)$$

where $n_{oTL} = \langle \delta a_{out\_m}^+\delta a_{out\_m}\rangle$ and $n_{oT} = \langle \delta b_{out\_n}^+\delta b_{out\_n}\rangle$ are the transmission line and transmon qubit output photon number, respectively. The criterion expressed in Eq. 15 reveals that for $\varepsilon_e > 1$ the modes become entangled [20]. In this criterion the phase sensitive factor has a critical role to create the entanglement between modes. The phase sensitive cross correlation is arisen due to the amplification-like interaction between the transmission line and transmon qubit modes. The entanglement metric in Eq. 15 emphasizes that the increase of the cross correlation factor leads to the entanglement between modes. In the following, we will try to demonstrate the effect of the photon number and also the phase sensitive cross correlation on the entanglement between modes. The simulation results (for the transmission line first mode

m = 1) are illustrated in Fig. 4 in which the transmission line outputs photon number (Fig. 4a), the transmon qubit output photon number (Fig. 4b), and the phase sensitive cross-correlation (Fig. 4c), and the entanglement between modes (Fig. 4d) are depicted. It is clearly shown in Fig. 4b that the number of the transmon qubit's photons are strongly affected due to the modes coupling from the transmission line around the JJ location (X = 0 in this simulation). As indicated with a dashed circle on Fig. 4c, it is shown that so close to X = 0 the magnitude of the phase sensitive cross-correlation is dramatically increased. This is the reason that the entanglement is created between modes neighboring the JJ location. From the quantum mechanical point of view, around the location indicated with the white dashed circle, the amount of the amplification-like interaction between modes are increased; this is the critical point to link the modes between transmission line and transmon qubit to create the entanglement. In other words, the maximum amount of the coupling occurs at a location so close to the JJ around X = 0. However, it should be noted from Fig. A1 that the maximum change of $\Delta E$ occurred around X = 0. It is clear that the more $\Delta E$ is affected in the system the more phase noise the transmon qubit is endured. In fact, this introduces a trade-off between the establishment of the entanglement and qubit incoherence. The latter result is one of the important goals achieved by this study.

Also, the contour maps illustrated in Fig. 4c and Fig. 4d demonstrates that the increase of $E_j/E_c$ is not a necessary condition to create the entanglement between modes. In contrast, one should concentrate on the detuning frequency that has a critical impact on the transmon qubit decay rate and accordingly the phase sensitive cross-correlation is influenced. For the current results the detuning frequency as $(\omega_{43} - \omega_{32})$ was applied. Other related detuning frequencies in the system that can be arisen are depicted in Fig. A3 in appendix A.

Table 1. Data for the simulation of the designed system [4, 5]

| Parameter | Value | Stand for |
|---|---|---|
| $C_g$ | 20 fF | Capacitance between transmission line and superconductors |
| $C_B$ | 40 fF | Capacitance between two superconductors |
| $C_J$ | 2 fF | Josephson junction capacitance |
| $C_0$ | 0.66 pF/m | transmission line capacitance |
| $L_0$ | 623 nH/m | transmission line inductance |
| $l_0$ | 12.7 mm | Length of transmission line |
| $C_{in}$ | 2 fF | Input capacitance |
| $E_c$ | 660 MHz | Charging energy |
| $E_J$ | 4.5 GHz | Josephson junction energy |

This suggests that for different detuning frequencies illustrated in Fig. A3 in Appendix A, the coupling between modes may change. One of the main factors affected because of the detuning frequency is the transmon decay rate, which is mathematically derived in Eq. 10. The mentioning point indicates that using different detuning frequency has a critical impact on the quantum feature of the system since the detuning frequency is essentially related to ΔE. In the designed system, it is supposed that the transmon qubit has four distinct energy levels. Thus, the detuning frequencies illustrated in Fig. A3 are derived based on the energy levels of the transmon qubit.

In contrast to the first mode of the transmission line that the entanglement is created between modes around X = 0, for the second mode due to the negligible coupling between them as shown in $E_2$ and $E_4$ in Fig. 2, the entanglement is completely lost. The simulation result is depicted in Fig. 5.

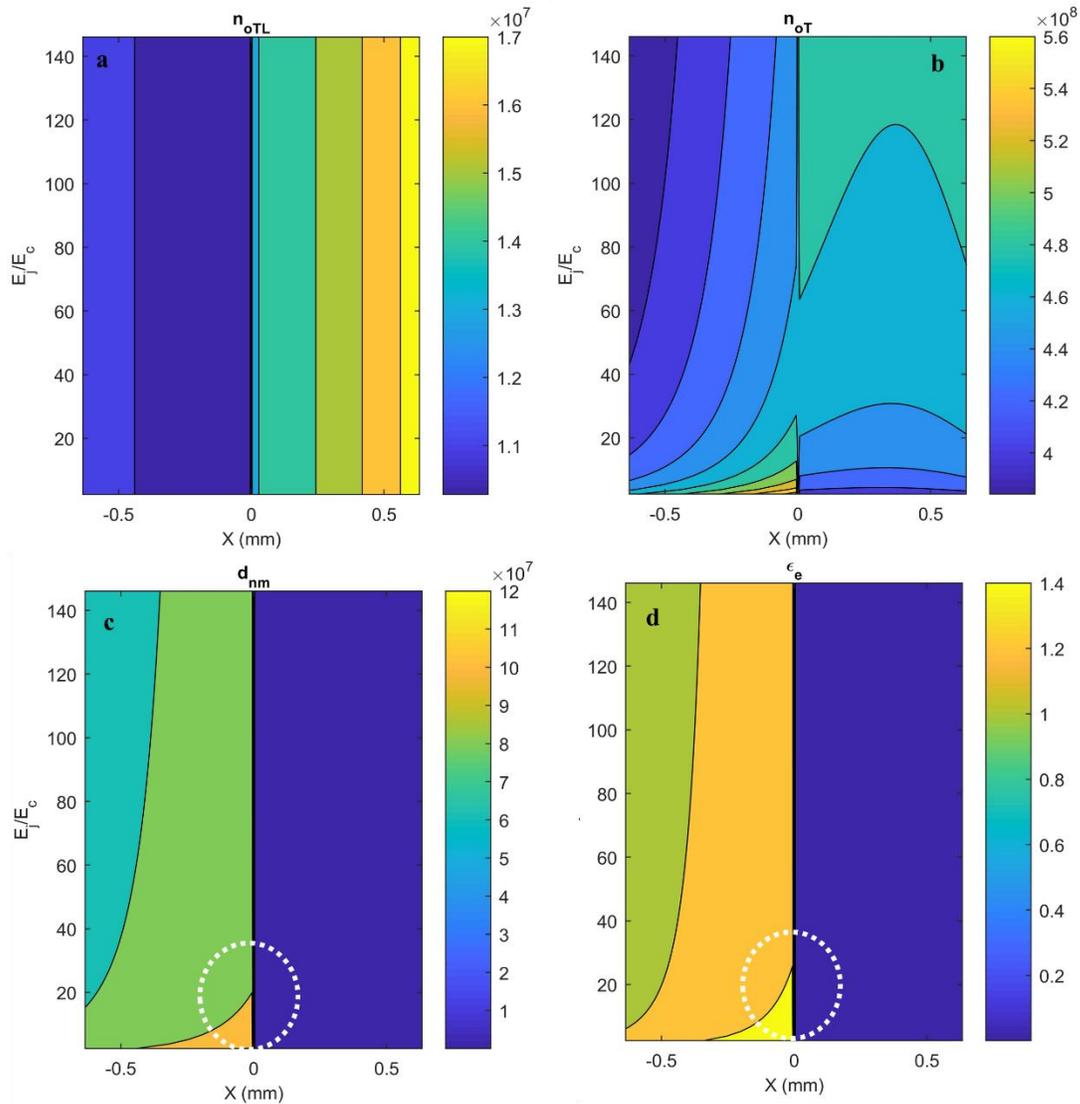

Fig. 4 a) Transmission line output number of photons, b) transmon qubit output number of photons, c) phase sensitive cross-correlation, d) Entanglement between modes vs. $E_J/E_c$ and X (mm); by considering the transmission line odd mode (m = 1).

Even through the transmission line even modes have a negligible overlapping with embedding qubit on it and this makes the modes completely become separable; but just at locations around X = 0 and on the left side, it seems that the phase sensitive cross correlation is strongly increased. Nonetheless, it is not enough to create the entanglement between modes. It consequently means that the coupling factor plays the critical role to generate the entanglement in the system.

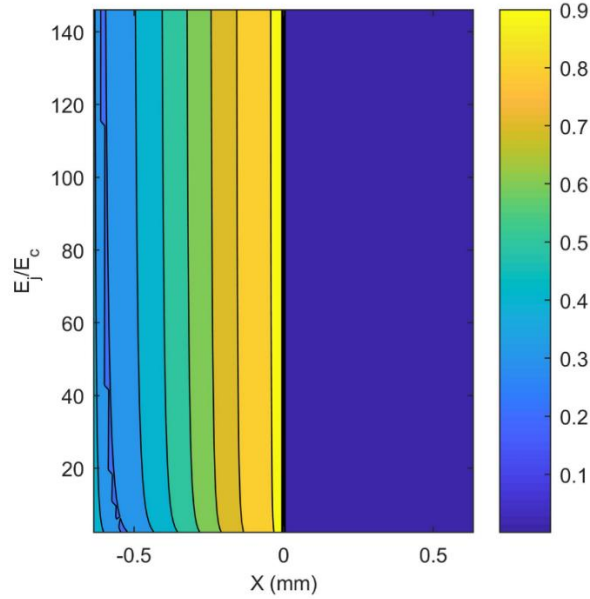

Fig. 5 Entanglement between modes vs. $E_J/E_c$ and X (mm); by considering the transmission line even mode (m = 2).

However, there are other factors that can affect the coupling between modes in the system. One of them is $C_g$, the capacitance between transmission line and the superconductors in the structure, by which $\alpha_m$, $\beta_m$, and $\gamma_m$ become affected. The effect of the mentioned capacitance on the entanglement between modes are depicted in Fig. A4 in Appendix A.

**Conclusions**

In this study, a system containing a transmon qubit embedding into a transmission line was designed and simulated. Initially, the system was analyzed quantum mechanically by which the related Lagrangian was derived and then the contributed quantum Hamiltonian was examined. Using the Hamiltonian of the system, the energy levels were calculated and then using them the different detuning frequencies to excite the transmon were estimated. Also, the anharmonicity of the coupling system is examined through which the minimum pulse duration was calculated and suggesting that the designed system could operate with the dephasing time around 100 psec. In the following, the study focused on establishing the entanglement between the transmission line and transmon qubit modes. For this reason, the dynamic equations of the motion of the system were derived and using them the photons number were calculated. Finally, employing the entanglement metric the entanglement between modes was calculated. As a main goal, it was shown that it is possible to create the entanglement between modes; in this way, some factors such as coupling between modes in the system and also $C_g$ could manipulate the entanglement between modes. This was the main aim that the system was designed for.

**References**

1. J. Koch, T. M. Yu, J. Gambetta, A. A. Houck, D. I. Schuster, J. Majer, A. Blais, M. H. Devoret, S/ M. Girvin, R. J. Schoelkopf, Charge-insensitive qubit design derived from the Cooper pair box, Phys. Rev. A, **76**, 042319, (2007).

2. J. Bourassa, F. Beaudoin, Jay M. Gambetta, and A. Blais, Josephson-junction-embedded transmission-line resonators: From Kerr medium to in-line transmon, Phys. Rev. A, **86**, 013814, (2012).

3. A. Blais, R.Sh. Huang, A. Wallraff, S. M. Girvin, and R. J. Schoelkopf, Cavity quantum electrodynamics for superconducting electrical circuits: An architecture for quantum computation, Phys. Rev. A, **69**, 062320, (2004).

4. A. P. Rodriguez, Circuit Quantum Electrodynamics with Transmon Qubits in the Ultrastrong Coupling Regime, Master in Quantum Science and technology, University of the Basque Country UPV/EHU, (2016).

5. Ch. K. Andersen, Theory and Design of Quantum devices in Circuit QED, PhD thesis, AARHUS University, (2016).

6. M. H. Devoret, R. J. Schoelkopf, Superconducting Circuits for Quantum Information: An Outlook, Science, **339**, 1169, (2013).

7. D. Vion, A. Aassime, A. Cottet, P. Joyez, H. Pothier, C. Urbina, D. Esteve, M. H. Devoret, Manipulating the Quantum State of an Electrical Circuit, Science, **296**, 886, (2002).

8. A. J. Berkley, H. Xu, R. C. Ramos, M. A. Gubrud, F. W. Strauch, P. R. Johnson, J. R. Anderson, A. J. Dragt, C. J. Lobb, F. C. Wellstood, Entangled Macroscopic Quantum States in Two Superconducting Qubits, Science, **300**, 1548, (2003).

9. Y. Shih, Entangled Photons, IEEE J. Sel. Top. Quantum Electron, **9**, 1455, (2003).

10. Y. Shih, Quantum Imaging, IEEE J. Sel. Top. Quantum Electron, **13**, 1016, (2007).

11. A. Salmanogli, D. Gokcen, H.S. Gecim, Entanglement Sustainability in Quantum Radar, IEEE J. Sel. Top. Quantum Electron, **26**, 1, (2020).

12. S. Barzanjeh, S. Guha, Ch. Weedbrook, D. Vitali, J. H. Shapiro, and S. Pirandola, Microwave Quantum Illumination, Phys. Rev. Lett. **114**, 080503, (2015).

13. A. Salmanogli, D. Gokcen, Design of quantum sensor to duplicate European Robins navigational system, Sensors and Actuators A: Physical, **322**, 112636, (2021).

14. A. Salmanogli, D. Gokcen, Entanglement Sustainability Improvement Using Optoelectronic Converter in Quantum Radar (Interferometric Object-Sensing), IEEE Sensors Journal, **21**, 9054, (2021).

15. A. Salmanogli, D. Gokcen, and H. S. Gecim, Entanglement of Optical and Microcavity Modes by Means of an Optoelectronic System, Phys Rev Applied, **11**, 024075, (2019).

16. S. Barzanjeh, S. Pirandola, D. Vitali, and J. M. Fink, Microwave quantum illumination using a digital receiver, Science Advances, **6**, 1, (2020).

17. A. Salmanogli, H. S. Gecim, Optical and Microcavity Modes Entanglement by Means of Plasmonic Opto-Mechanical System, IEEE J. Sel. Top. Quantum Electron, **26**, 1016, (2020).

18. A. Salmanogli, Entanglement Generation using Transistor Nonlinearity in Low Noise Amplifier, under review by Physical Review Applied, (2021).

19. A. Salmanogli, Design of a Quantum Radar System with Sustainable Entanglement, PhD thesis, Hacettepe University, (2021).

20. S. Barzanjeh, S. Guha, Ch. Weedbrook, D. Vitali, J. H. Shapiro, and S. Pirandola, Microwave Quantum Illumination, Phys. Rev. Lett. **114**, 080503, (2015).

21. A. Salmanogli, Quantum analysis of plasmonic coupling between quantum dots and nanoparticles, Phys. Rev. A, **94**, 043819, (2016).

22. A. Salmanogli, Modification of a plasmonic nanoparticle lifetime by coupled quantum dots, Phys. Rev. A, **100**, 013817, (2019).

**Appendix A:**

In Fig. A1, difference energy levels of the transmon qubit is illustrated. The graphs actually show the behavior of ΔE by which one can figure out where the maximum noise can be applied.

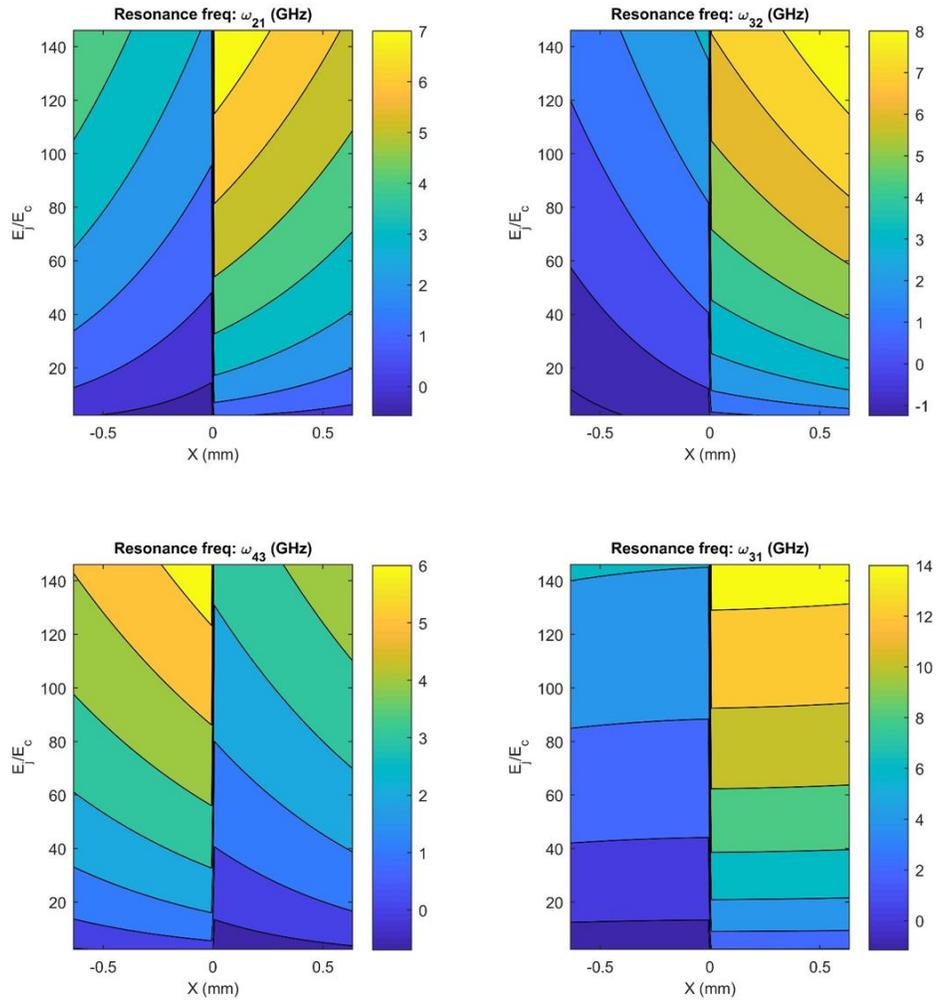

Fig. A1 Different energy levels of the transmon (GHz) vs. $E_J/E_c$ and X (mm)

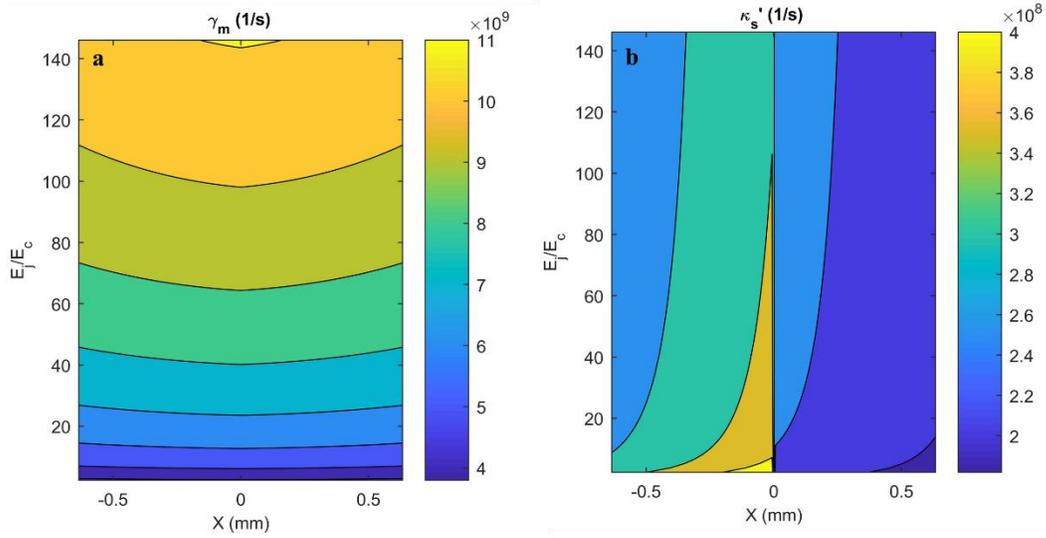

Fig. A2 $\gamma_m$ and $\kappa_s$' vs. $E_J/E_c$ and X (mm)

In Fig. A2 the coupling system decay rate and also the transmon qubit's modified decay rate are illustrated. As one can expect, the modified decay rate became maximized around the Josephson Junction location.

Finally, the different detuning frequencies for the designed system is illustrated in Fig. A3. The results depicted in Fig. A3 are comparable with the results that have been derived for the transmon qubit detuning frequencies in [2].

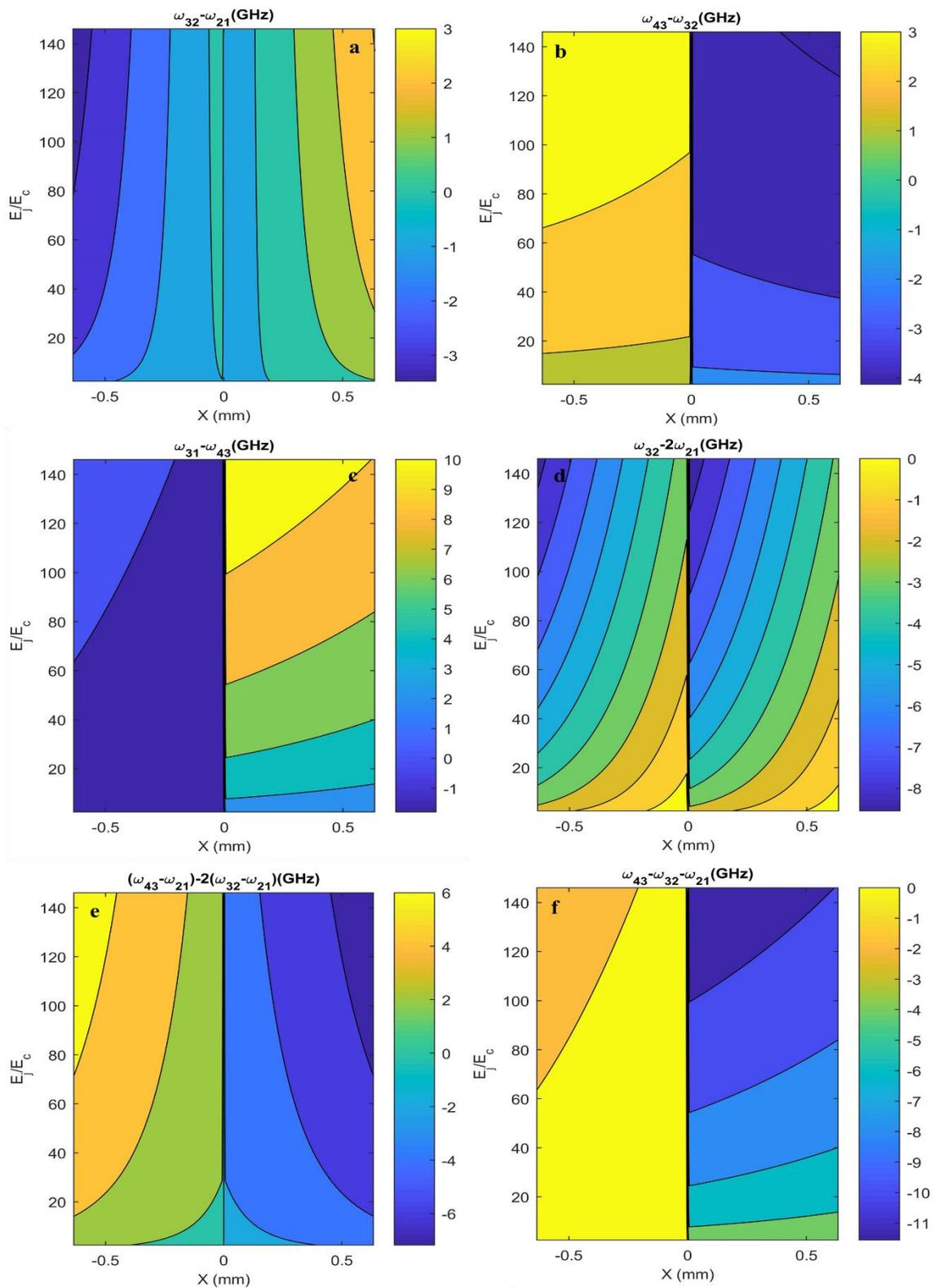

Fig. A3 Different detuning frequencies vs. $E_J/E_c$ and X (mm)

In Fig. A4 the effect of the mentioned capacitance on the entanglement between modes are depicted. It is shown that increasing $C_g$ leads to decrease of the entanglement between modes in the system. It is because increasing $C_g$ leads to decreasing $\alpha_m$ and $\beta_m$ by which the coupling between modes is decrease and that is the reason by which $d_{nm}$ in the coupled system is decreased.

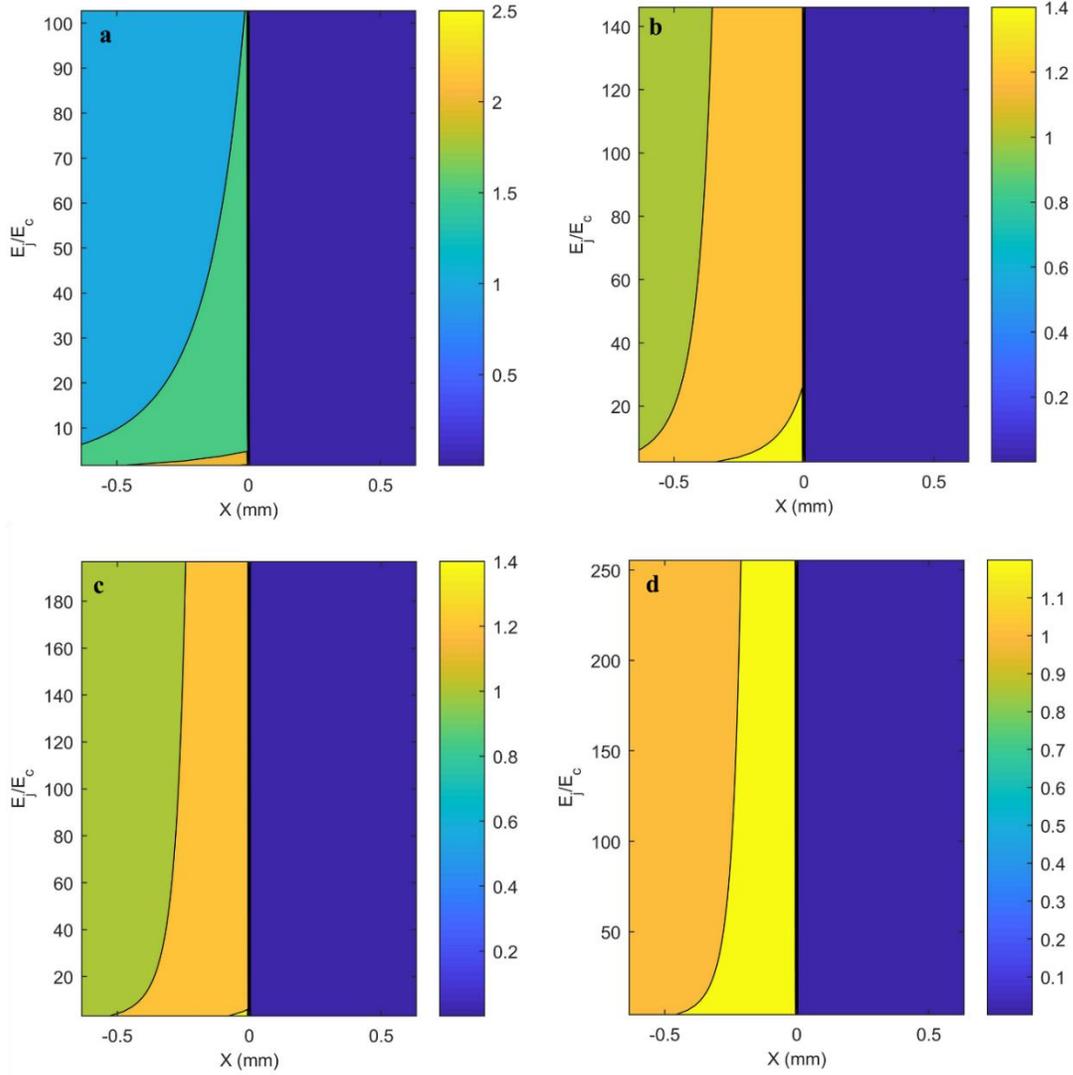

Fig. A4 Entanglement between modes vs. $E_J/E_c$ and X (mm), a) $C_g = 10$ fF, b) $C_g = 20$ fF, c) $C_g = 30$ fF, d) $C_g = 40$ fF, transmission line first mode m = 1.

**Appendix B:**

Dynamics equation of motion for the strong field:

$$0 = -\left(j\Delta_m + \frac{\kappa_m}{2}\right)A_m + j\gamma_m\left(B_n - B_n^*\right)$$
$$0 = -\left(j\Delta_n + \frac{\kappa_n}{2}\right)B_n + j\gamma_m\left(A_m - A_m^*\right) + j\gamma_N\left(B_n + B_n^+\right)^3 \tag{B1}$$

Solving equations expressed in Eq. B1 leads to determine $A_m$ and $B_n$ as the transmission line and transmon qubit strong fields, respectively. The strong field points are substituted in Eq. 11 to find the transmission line and transmon qubit fluctuation modes.

Using Eq. 12 of the main paper and making some definitions such as $A_0 = j\Delta_m + \kappa_m/2$ and $B_0 = j\Delta_n + \kappa_n/2$ and, the transmission line number of photons are calculated as:

$$n_{TL} \equiv <\delta a_m^+ \delta a_m> = \frac{\gamma_m^2}{|A_0|^2}\left(<\delta b_n^+ \delta b_n> + <\delta b_n \delta b_n^+>\right) + \frac{1}{|A_0|^2}<\delta a_{in}^+ \delta a_{in}> \tag{B2}$$

The number of photons of the transmon is calculated using:

$$n_T \equiv <\delta b_n^+ \delta b_n> = \frac{\gamma_m^2}{|B_0|^2}\left(<\delta a_m^+ \delta a_m> + <\delta a_m \delta a_m^+>\right) + \frac{\gamma_N^2}{|B_0|^2}\left(<\delta b_n \delta b_n^+>\right) + \frac{1}{|B_0|^2}<\delta b_{in}^+ \delta b_{in}> \tag{B3}$$

The Eq. B3 is re-expressed as:

$$n_T\left(1 - \frac{\gamma_N^2}{|B_0|^2}\right) = \frac{\gamma_m^2}{|B_0|^2}\left(2n_{TL} + 1\right) + \frac{\gamma_N^2}{|B_0|^2} + \frac{1}{|B_0|^2}<n_{inb}> \tag{B4}$$

where $n_{inb} = <\delta b_{in}^+ \delta b_{in}>$, $n_{ina} = <\delta a_{in}^+ \delta a_{in}>$ and also we define $B_1 = 1 - \gamma_N^2/|B_0|^2$.